\documentclass[conference]{IEEEtran}
\IEEEoverridecommandlockouts

\usepackage{cite}
\usepackage{amsmath,amssymb,amsfonts}
\usepackage{graphicx}
\usepackage{textcomp}
\usepackage{xcolor}
\usepackage{algorithm}
\usepackage{algpseudocode}
\usepackage{booktabs}
\usepackage{multirow}
\usepackage{hyperref} 
\usepackage{soul}
\usepackage{enumitem}

\def\BibTeX{{\rm B\kern-.05em{\sc i\kern-.025em b}\kern-.08em
    T\kern-.1667em\lower.7ex\hbox{E}\kern-.125emX}}
\usepackage{url}

\begin{document}

\title{A2RAG: Adaptive Agentic Graph Retrieval for Cost-Aware and Reliable Reasoning}


\author{
Jiate Liu\textsuperscript{1}, 
Zebin Chen\textsuperscript{1}, 
Shaobo Qiao\textsuperscript{2}, 
Mingchen Ju\textsuperscript{2}, 
Danting Zhang, 
Bocheng Han\textsuperscript{1},
Shuyue Yu\textsuperscript{3,4},\\
Xin Shu\textsuperscript{1,4},
Jinglin Wu\textsuperscript{3,4},
Dong Wen\textsuperscript{1},
Xin Cao\textsuperscript{1},
Guanfeng Liu\textsuperscript{5}
and Zhengyi Yang\textsuperscript{1,*}\thanks{*Zhengyi Yang is the corresponding author.} \\

\fontsize{10}{10}
\selectfont\itshape
\textsuperscript{1}\textit{University of New South Wales}, 
\textsuperscript{2}\textit{Euler AI}, 
\textsuperscript{3}\textit{Sigma Trading Management}, 
\textsuperscript{4}\textit{Eigenflow AI}, 
\textsuperscript{5}\textit{Macquarie University}, 
\\

\fontsize{9}{9} \selectfont\ttfamily\upshape
\textsuperscript{1}\{jiate.liu, zebin.chen1, bocheng.han, xin.shu7, dong.wen, xin.cao, zhengyi.yang\}@unsw.edu.au; \\
\textsuperscript{2}\{shaobo.qiao, mingchen.ju\}@eulerai.au; 
\textsuperscript{3}\{carol.yu, anthony.wu\}@sigmatm.com.au; \\
\textsuperscript{5}  guanfeng.liu@mq.edu.au; 
dantingxing1994@outlook.com %
}

\maketitle

\begin{abstract}

Graph Retrieval-Augmented Generation (Graph\-RAG) enhances multi\-hop question answering by organizing corpora into knowledge graphs and routing evidence through relational structure. However, practical deployments face two persistent bottlenecks: (i) mixed-difficulty workloads where one-size-fits-all retrieval either wastes cost on easy queries or fails on hard multi\-hop cases, and (ii) extraction loss, where graph abstraction omits fine-grained qualifiers that remain only in source text.
We present \textbf{A2RAG}, an \emph{adaptive-and-agentic} GraphRAG framework for cost-aware and reliable reasoning. 
A2RAG couples an adaptive controller that verifies evidence sufficiency and triggers targeted refinement only when necessary, with an agentic retriever that progressively escalates retrieval effort and maps graph signals back to provenance text to remain robust under extraction loss and incomplete graphs.
Experiments on HotpotQA and 2Wiki\-MultiHopQA demonstrate that A2RAG achieves \textbf{+9.9/+11.8 absolute gains in Recall@2}, while cutting token consumption and end-to-end latency by about \textbf{50\%} relative to iterative multi\-hop baselines.
\end{abstract}


\section{Introduction}
\label{sec:introduction}

Recently, Large Language Models (LLMs) have demonstrated transformative capabilities across diverse domains~\cite{chowdhery2023palm}, 
including natural language understanding~\cite{chowdhery2023palm,lai2025graphy}, 
code generation~\cite{wang-etal-2023-codet5}, 
multi-modal retrieval~\cite{pmlr-v139-radford21a,tang2025tabular,wu2024experimental}, 
and complex reasoning~\cite{wei2022chainofthought,yao2023react,chen2023machine}.

However, in critical sectors such as finance, healthcare, and law, their practical utility remains fundamentally limited by a lack of intrinsic grounding. 
This limitation frequently leads to hallucinations---the generation of statements that are linguistically plausible yet unsupported by reliable evidence~\cite{huang2023hallucination_survey}. 
For example, in financial risk monitoring or legal compliance, where every decision must be traceable and auditable, such hallucinations pose unacceptable operational risks.

To mitigate these issues, Retrieval-Augmented Generation (RAG) has emerged as a standard solution by augmenting LLMs with an external corpus and conditioning generation on retrieved evidence~\cite{lewis2020rag,gao2023rag_survey}. 
Despite its success, conventional RAG systems typically rely on dense retrieval and similarity-based search to fetch isolated text chunks~\cite{karpukhin2020dpr,izacard2021fid}. 
This paradigm often overlooks latent structural relations among entities and struggles to synthesize information distributed across multiple documents, leading to \emph{context fragmentation}, where the LLM is provided with noisy and weakly connected evidence snippets.

Graph Retrieval-Augmented Generation (GraphRAG) addresses this limitation by explicitly modeling entities and their relations as a knowledge graph, thereby enabling retrieval and reasoning over structured dependencies~\cite{edge2024local2global_graphrag,peng2024graphrag_survey}. By converting unstructured text into a structured graph representation, GraphRAG can navigate semantic dependencies and compose multi-hop evidence through graph traversal, bridging informational gaps that are difficult to capture with flat, vector-based retrieval. Recent work further explores graph-guided reasoning paradigms that integrate structured reasoning paths into LLM inference~\cite{jin2024graphcot,huan2025glm}.

\noindent\textbf{Motivation and Challenges.} 
Despite its promise, deploying GraphRAG in practical and production-scale environments exposes two fundamental bottlenecks that existing systems have yet to adequately address.

\noindent\textit{\ul{Challenge 1: One-size-fits-all retrieval under mixed-difficulty workloads.}}
Existing GraphRAG systems typically adopt a fixed retrieval strategy: some~\cite{edge2024local2global_graphrag,peng2024graphrag_survey}
prioritize graph-wide \emph{global} operations (e.g., community-summary-based global search in Microsoft GraphRAG),
while others~\cite{guo2025lightrag} are designed to remain within a cheap local neighborhood.
However, our empirical analysis on production-level queries---collected from real-world user interactions on a foreign exchange (FX) trading platform (\emph{LP Margin})~\cite{wang2025aefa}, together with standard multi-hop benchmarks~\cite{yang2018hotpotqa,ho2020constructing}---reveals a pronounced workload skew: approximately 60\% of queries can be answered using simple, low-cost retrieval, whereas the remaining 40\% require genuinely complex, multi-hop evidence composition.

This mismatch leads to two clear failure modes. When a system \emph{always} applies complex retrieval, it incurs unnecessary latency and token overhead on the majority of easy queries, and may even introduce additional noise by over-expanding the context. Conversely, when a system \emph{always} relies on simple retrieval, it fails precisely on the hard-tail cases where evidence is distributed and must be composed across multiple hops. The core difficulty is that this “easy vs.\ hard” distinction is not reliably detectable a priori: surface-level signals such as keyword overlap or query length are often insufficient to predict whether multi-hop reasoning will be required.

At a deeper level, this reflects a fundamental \emph{decision-making gap}: current GraphRAG systems lack an adaptive mechanism that can \emph{assess evidence sufficiency during retrieval} and decide whether to terminate early or escalate retrieval effort under a budget. Without such a progressive, evidence-aware policy, systems are forced into a one-size-fits-all routine that results in either systematic waste or systematic failure. While some prior work has explored adaptive retrieval via question complexity estimation or self-reflection, these approaches are largely text-centric and do not provide a graph-specific design that progressively escalates retrieval while preserving provenance and controllability~\cite{jeong2024adaptive_rag,asai2023selfrag,wang2025adaptive_rag_conversational}.

\noindent\textit{\ul{Challenge 2: Vulnerability to extraction loss and knowledge incompleteness.}}
In practice, a knowledge graph constructed from text typically captures the \emph{coarse} ``who-is-related-to-what'' structure, but it often misses the \emph{fine-grained qualifiers} that determine correctness in real decision-making. We repeatedly observe this phenomenon in production-level corpora: many queries depend on conditions, numerical thresholds, temporal qualifiers, and exceptions, yet these details are frequently not represented explicitly once the corpus is converted into a graph for retrieval~\cite{simple_is_effective_kgrag_2024}. As a result, a system that relies solely on the graph may return answers that appear structurally plausible but are imprecise or even incorrect when the missing qualifiers are critical.

A key reason is that typical extraction pipelines are optimized to produce clean subject--relation--object triples, which incentivizes dropping qualifiers that do not fit naturally into a single edge label. We refer to this systematic omission as \emph{extraction loss}: the resulting graph becomes a simplified projection of the corpus, while crucial details remain only in the original text. Moreover, real-world data is inherently noisy: entity mentions may overlap, relations are often implicit, and extraction errors can fragment the graph, leading to incomplete and uneven coverage.

For example, consider a financial compliance query regarding trade limits. A source document might state: ``Bank A is permitted to execute high-leverage trades in Region X, provided the daily volatility remains below 2.0\%.'' An extraction pipeline may retain only the triple \texttt{(Bank A, permitted\_in, Region X)}, discarding the critical conditional constraint on volatility. 
As another example from an Foreign Exchange (FX) risk corpus, the same liquidity provider may appear as ``CFH'', ``CFH Clearing'', or ``[GUI] CFH'' across logs and reports. If the extractor fails to normalize these aliases, the graph will contain multiple disconnected nodes for the same real-world entity, causing relevant facts to be split across components and preventing retrieval from following the intended evidence chain.

Simply ``fixing the graph'' by making extraction increasingly sophisticated is often impractical: it requires case-specific prompts and frequent re-extraction, which becomes prohibitively expensive to maintain in dynamic environments. Importantly, our observation is that even when fine-grained details are missing, the graph's \emph{connectivity structure} is often largely correct. This suggests that the appropriate role of the graph is to serve as a \emph{navigational map} to locate relevant regions of the corpus, while the system must ultimately recover precise and auditable evidence from the original source text. The core challenge, therefore, is to design retrieval mechanisms that are robust to imperfect and detail-poor graphs, using structure for routing but relying on provenance text for final, high-precision answering~\cite{simple_is_effective_kgrag_2024}.

\noindent\textbf{Our Solution and Contributions.} 
To address these challenges, we present A2RAG (\textbf{A}daptive \textbf{A}gentic Graph \textbf{R}etrieval-\textbf{A}ugmented \textbf{G}eneration), a unified framework that decouples \emph{answer-level reliability control} from \emph{retrieval-level progressive evidence acquisition}. We formulate retrieval as a dynamic, cost-aware process governed by two complementary layers:

\noindent\textit{\ul{Adaptive Control Loop.}} 
A reliability-oriented closed-loop mechanism that verifies the \emph{final} answer against retrieved provenance via \emph{Triple-Check} (relevance, grounding, and adequacy). Rather than micromanaging individual retrieval operators, the controller diagnoses failure modes and triggers \emph{failure-aware query rewriting} with bounded retries, ensuring that subsequent retrieval attempts are better targeted instead of repeatedly executing the same ineffective routine. This extends the spirit of adaptive retrieval to graph-augmented settings~\cite{jeong2024adaptive_rag,asai2023selfrag}.

\noindent\textit{\ul{Agentic Retriever.}} 
A stateful agent that executes a progressive, local-first retrieval policy with \emph{stage-wise evidence sufficiency checks}. It early-stops on easy queries after inexpensive local expansion, escalates to bounded bridge discovery when multi-hop connectors are required, and employs a global Personalized PageRank (PPR)-guided fallback as a last resort~\cite{page1999pagerank,haveliwala2002topicsensitive}. Crucially, it maps high-relevance graph regions back to provenance text chunks to recover precise conditions, numerical constraints, and qualifiers that may be missing from imperfect graphs, using the graph primarily as a navigational scaffold rather than as a complete semantic store~\cite{simple_is_effective_kgrag_2024}.

In summary, our contributions are threefold:

\begin{itemize} [leftmargin=*]
  \item \textbf{Evidence-Sufficiency Adaptive Control Loop.} We introduce a closed-loop controller that verifies evidence at the answer level and triggers refinement or escalation only when evidence sufficiency is not met, mitigating the cost--capability mismatch of static retrieval.
  
  \item \textbf{Agentic Graph Retrieval with Provenance Recovery.} We propose a progressive retriever with an explicit escalation policy and a provenance map-back mechanism, enhancing robustness to extraction loss compared to graph-only baselines.
  
  \item \textbf{Extensive Empirical Evaluation.} We evaluate A2RAG on two public datasets and one production dataset. On the public benchmarks HotpotQA and 2WikiMultiHopQA~\cite{yang2018hotpotqa,ho2020constructing}, A2RAG achieves absolute gains of 9.9\% and 11.8\% in Recall@2, respectively, while reducing token usage and latency by approximately 50\% compared to iterative baselines such as IRCoT~\cite{trivedi2023ircot}.

\end{itemize}

\section{Background and Related Work}
\label{sec:related_work}

\subsection{Problem Definition}
\label{sec:problem_definition}

\noindent\textbf{Setting}
We study knowledge-intensive question answering where a system must answer a
natural-language query using \emph{verifiable} evidence, and the required facts
may be distributed across multiple sources (multi-hop).

\noindent\textbf{Notation}
Let the input be a query $q$ and a text corpus
$\mathcal{D}=\{d_1,\ldots,d_N\}$ consisting of provenance passages/chunks.
We assume an offline-constructed knowledge graph
$\mathcal{G}=(\mathcal{V},\mathcal{E}_{\mathcal{G}})$ built from $\mathcal{D}$
via an information extraction pipeline, where $\mathcal{V}$ denotes entity nodes
and $\mathcal{E}_{\mathcal{G}}$ denotes relation edges.
We further assume an offline map-back function $\pi:\mathcal{V}\to 2^{\mathcal{D}}$
that returns the set of provenance chunks in $\mathcal{D}$ associated with a node
(e.g., mentioning or defining the entity).

\noindent\textbf{Goal}
The system outputs an answer $a$ together with an evidence set
$\mathcal{E}\subseteq\mathcal{D}$ that grounds $a$, or returns an abstention
signal when sufficient evidence cannot be obtained under a bounded retrieval
budget.

\noindent\textbf{Motivation}
Automatically constructed graphs summarize text into entities and relations and
may omit fine-grained details (e.g., numbers, temporal qualifiers, constraints),
an effect we refer to as \emph{extraction loss}
\cite{edge2024local2global_graphrag,simple_is_effective_kgrag_2024}.
In deployment, retrieval must therefore balance (i) efficiency versus coverage
for hard multi-hop queries, (ii) maintenance overhead under corpus updates, and
(iii) robustness via faithful provenance recovery from source text when graph
abstractions are insufficient.

\subsection{Related Methods}
\label{sec:existing_methods}

Prior work on retrieval for knowledge-intensive multi-hop QA spans graph-augmented
pipelines, iterative multi-hop text retrieval, and adaptive/agentic control
mechanisms \cite{peng2024graphrag_survey}.

\noindent\textbf{Local graph retrieval.}
Lightweight GraphRAG variants emphasize efficiency by anchoring retrieval on
query-aligned entities and restricting expansion to local neighborhoods, often
combined with lexical matching. LightRAG is a representative example that
targets low-latency, simple deployment while still leveraging graph structure for
routing \cite{guo2025lightrag, peng2024graphrag_survey}. A common limitation of
local-first access is that it can still struggle with harder multi-hop queries
that require discovering non-local connectors, motivating mechanisms that
escalate beyond immediate neighborhoods when local evidence is insufficient
\cite{guo2025lightrag,peng2024graphrag_survey}.

\noindent\textbf{Global summarization and indexing.}
At the other extreme, Microsoft GraphRAG constructs global structure by detecting
graph communities (e.g., Leiden \cite{traag2019leiden}) and generating
hierarchical natural-language summaries to support query-focused summarization
and corpus-level sense-making \cite{edge2024local2global_graphrag,peng2024graphrag_survey}.
While this paradigm can improve coverage for broad queries, it typically requires
substantial offline computation, and rebuilding or refreshing graph-derived
indices/summaries as the corpus evolves can incur non-trivial maintenance
overhead \cite{edge2024local2global_graphrag,peng2024graphrag_survey}.

\noindent\textbf{Iterative text retrieval for multi-hop reasoning.}
A line of work improves multi-hop coverage by interleaving retrieval with
step-by-step reasoning in flat text space. IRCoT exemplifies this strategy by
alternating intermediate reasoning with follow-up retrieval to gather distributed
evidence across passages \cite{trivedi2023ircot}. Such text-only iterative
retrieval often requires multiple rounds of querying and context accumulation,
which may increase runtime cost and amplify noisy evidence compared to
structure-guided routing \cite{trivedi2023ircot,gao2023rag_survey}.

\noindent\textbf{Adaptive and agentic retrieval control.}
Adaptive retrieval frameworks introduce dynamic control to allocate retrieval
effort based on query difficulty, with actions such as deciding when to retrieve,
reflect/critique retrieved content or intermediate answers, and when to stop or
skip retrieval \cite{jeong2024adaptive_rag,asai2023selfrag,wang2025adaptive_rag_conversational}.
Most of these designs are developed primarily for flat text retrieval and do not
explicitly model graph-native escalation operators (e.g., local neighborhood
expansion versus bridge discovery versus global diffusion), which are central to
GraphRAG-style pipelines \cite{jeong2024adaptive_rag,peng2024graphrag_survey}.

\noindent\textbf{Robustness to structural abstraction.}
Graph-based retrieval relies on offline information extraction, and the resulting
structural abstraction can omit fine-grained details that remain in the original
text (e.g., numbers, temporal qualifiers, exceptions). This motivates provenance
recovery mechanisms that map structure-guided signals back to source passages for
faithful grounding \cite{simple_is_effective_kgrag_2024,peng2024graphrag_survey}.

\noindent\textbf{Takeaway}
Overall, global GraphRAG improves coverage but increases build and refresh cost
\cite{edge2024local2global_graphrag}, local graph retrieval is efficient but can
miss non-local connectors \cite{guo2025lightrag}, iterative text retrieval
improves multi-hop coverage at the expense of multiple rounds of retrieval
\cite{trivedi2023ircot}, and adaptive agents largely operate over flat text
actions rather than graph-native escalation \cite{jeong2024adaptive_rag,asai2023selfrag,wang2025adaptive_rag_conversational}.
These gaps motivate a unified framework that supports cost-aware escalation over
graph-native operations while recovering faithful provenance from source text
under structural abstraction \cite{simple_is_effective_kgrag_2024,peng2024graphrag_survey}.

\section{A2RAG Framework}

\subsection{Overview}
\label{sec:overview}

We present A2RAG, a unified framework that decouples \emph{adaptive control} from
\emph{agentic retrieval} to enable cost-aware and reliable evidence acquisition
for knowledge-intensive multi-hop QA.

\begin{figure}[t]
  \centering
  \includegraphics[width=\linewidth]{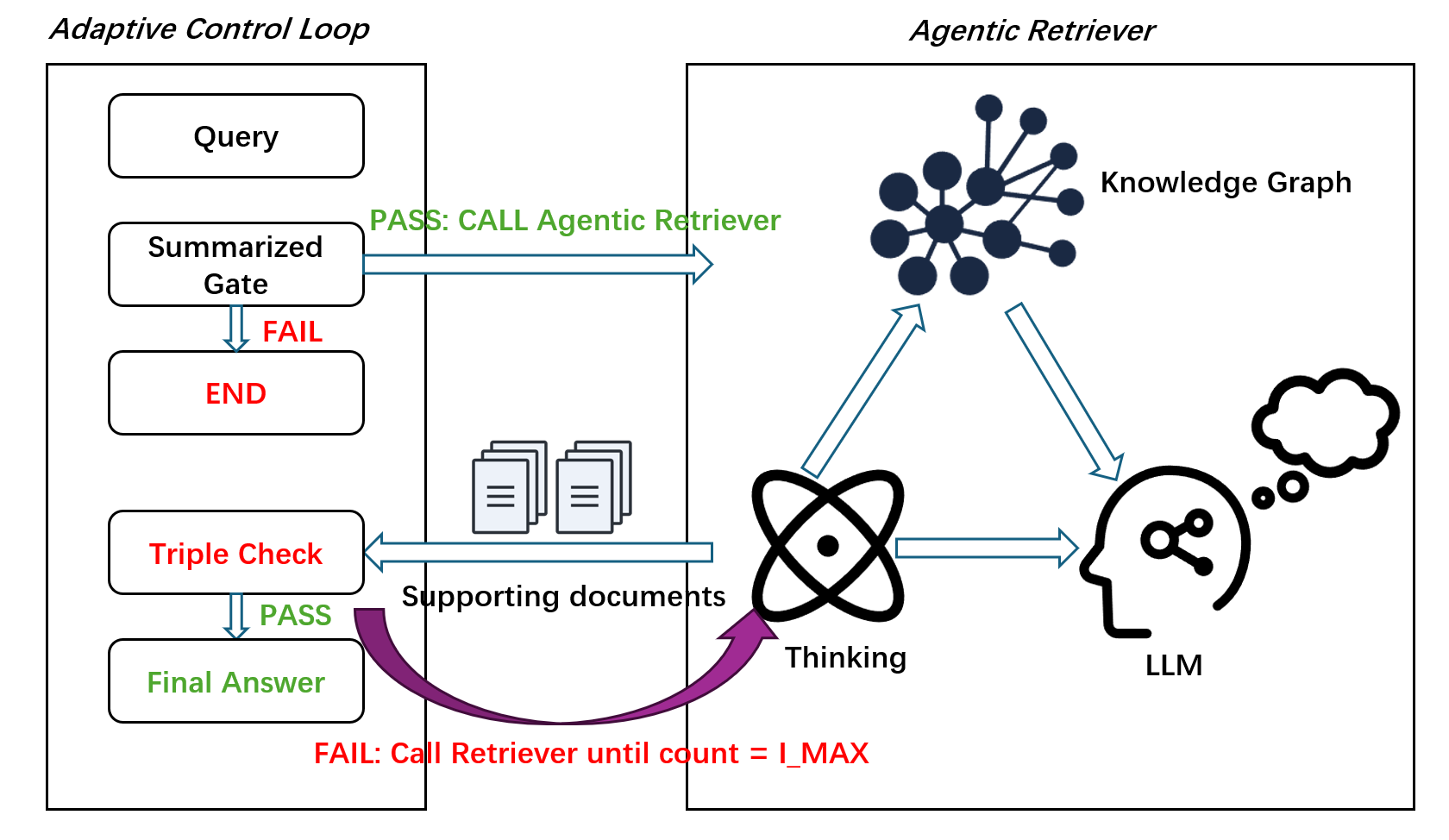}
  \caption{A2RAG Framework Overview}
  \label{fig:framework_overview}
\end{figure}

\noindent\textbf{Two-Layer Architecture.}
A2RAG consists of two tightly coupled components (Figure~\ref{fig:framework_overview}).

\textbf{(1) Adaptive Control Loop.}
The controller governs the retrieval lifecycle at a global level. It decides
whether to retrieve via lightweight gating, generates a candidate answer, and
performs answer-level verification for relevance, grounding, and query
resolution. When verification fails, it triggers failure-aware query rewriting
and bounded retries under a given budget. Human-in-the-loop validation is
supported as an optional extension for safe knowledge base evolution.

\textbf{(2) Agentic Retriever.}
Once retrieval is invoked, the retriever acts as a stateful agent that performs
progressive evidence discovery with a local-first, escalation-on-demand policy.
It performs stage-wise evidence sufficiency checks after each retrieval stage to
decide whether to escalate for additional evidence (sufficiency for escalation
decisions rather than certification of answer correctness). When the graph is
sparse or suffers from extraction loss, the retriever recovers fine-grained
provenance by mapping high-scoring graph nodes or regions back to source text
chunks in $\mathcal{D}$.

\noindent\textbf{Component interaction.}
The controller invokes the retriever with the current query state (including any
rewritten query), and the retriever returns provenance evidence for answering.
The controller then generates an answer and verifies it; upon failure, it
rewrites the query and re-invokes retrieval within a bounded budget. This
separation of concerns enables reliable and cost-aware answering without
committing to a fixed retrieval pipeline.

We now describe each component in detail, beginning with the adaptive control
loop (Section~\ref{sec:control_loop}) and followed by the agentic retriever
(Section~\ref{sec:agentic_retriever}).

\subsection{Adaptive Control Loop}
\label{sec:control_loop}

\begin{figure}[t]
  \centering
  \includegraphics[width=\linewidth]{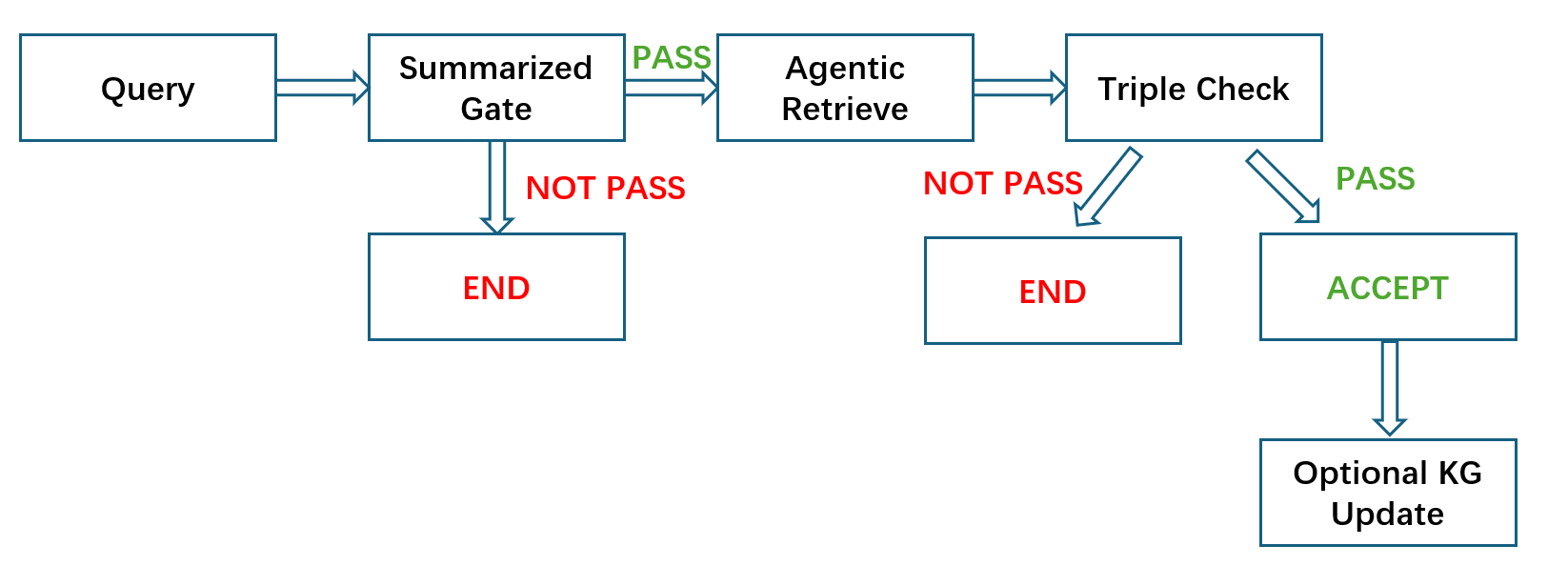}
  \caption{Adaptive Control Loop}
  \label{fig:control_loop}
\end{figure}

The adaptive control loop orchestrates retrieval and answering under practical
constraints. It (i) filters out out-of-scope queries before retrieval, (ii)
verifies answer correctness against provenance evidence, and (iii) revises the
query with bounded retries when evidence is misaligned or insufficient.
Figure~\ref{fig:control_loop} summarizes the control flow.

\subsubsection{Entry: Summarized-KB Gating}
\label{sec:kb_gating}

In real deployments, many queries are out-of-scope or only weakly related to the
indexed corpus. Running retrieval and large-model inference for such queries
wastes budget and may introduce spurious evidence. We therefore apply a
lightweight gate that estimates corpus coverage before invoking retrieval.

\noindent\textbf{Offline summaries.}
For each document $d_j\in\mathcal{D}$, we precompute a short summary $s_j$ during
indexing and collect all summaries into $\mathcal{S}=\{s_j\}_{j=1}^{N}$, where
$N:=|\mathcal{D}|$.

\noindent\textbf{Gate decision.}
Given a query $q$, we compute a similarity score $\gamma_j(q)\in[0,1]$ between
$q$ and each summary $s_j$ (e.g., cosine similarity between dense embeddings,
optionally clipped to $[0,1]$). We define
\begin{equation}
\mathrm{Gate}(q) := \mathbb{1}\!\left\{\max_{1\le j\le N} \gamma_j(q) \ge \tau_g\right\}\in\{0,1\},
\label{eq:gating_simple}
\end{equation}
where $\tau_g\in[0,1]$ is a threshold and $\mathbb{1}\{\cdot\}$ is the indicator
function. If $\mathrm{Gate}(q)=0$, the system returns \textsc{Abstain} (or falls
back to a lightweight non-retrieval mode); otherwise, the controller invokes the
agentic retriever.

\subsubsection{Exit: Verified Answering via Triple-Check}
\label{sec:triple_check}

Retrieval alone does not guarantee faithful answers. Retrieved passages may be
off-topic, the model may generate unsupported claims, or the answer may fail to
resolve the query. We therefore enforce a verification layer that checks three
complementary properties.

Let $\mathcal{E}\subseteq\mathcal{D}$ denote the retrieved provenance evidence
(i.e., text chunks from the corpus), and let $a\in\text{Text}$ be a candidate
answer generated by a language model conditioned on $(q,\mathcal{E})$. We define
three binary validators
\begin{align}
V_{\mathrm{rel}}(q,\mathcal{E}) &\in \{0,1\}, && \text{evidence relevance}, \label{eq:v_rel_simple}\\
V_{\mathrm{grd}}(a,\mathcal{E}) &\in \{0,1\}, && \text{answer grounded in evidence}, \label{eq:v_grd_simple}\\
V_{\mathrm{ans}}(q,a) &\in \{0,1\}, && \text{query resolution / adequacy}. \label{eq:v_ans_simple}
\end{align}
The overall predicate is their conjunction
\begin{equation}
\mathrm{TripleCheck}(q,a,\mathcal{E}) :=
V_{\mathrm{rel}}(q,\mathcal{E}) \land V_{\mathrm{grd}}(a,\mathcal{E}) \land V_{\mathrm{ans}}(q,a),
\label{eq:triplecheck_simple}
\end{equation}
where $\land$ denotes logical AND. Each validator can be instantiated using an
NLI model or a prompted LLM-based binary classifier. The controller accepts $a$
iff $\mathrm{TripleCheck}(q,a,\mathcal{E})=1$; otherwise it proceeds to
iterative refinement.

\subsubsection{Iteration: Failure-Aware Rewrite and Bounded Retry}
\label{sec:query_rewriting}

When verification fails, simply re-running retrieval with the same query often
repeats the same mistakes (e.g., retrieving the same neighborhood or the same
off-topic passages). We instead rewrite the query based on the failure mode and
retry within a bounded budget.

\noindent\textbf{Failure type and rewrite.}
At controller iteration $i$, let $(q^{(i)}, a^{(i)}, \mathcal{E}^{(i)})$ denote
the current query, candidate answer, and evidence, with $q^{(0)}:=q$. We define
a failure type $\mathrm{type}^{(i)}\in\{\mathrm{rel},\mathrm{grd},\mathrm{ans}\}$
as the first violated condition among
$V_{\mathrm{rel}}(q^{(i)},\mathcal{E}^{(i)})$,
$V_{\mathrm{grd}}(a^{(i)},\mathcal{E}^{(i)})$, and
$V_{\mathrm{ans}}(q^{(i)},a^{(i)})$.
We then rewrite the query with a type-conditioned function
\begin{equation}
q^{(i+1)} :=
\mathrm{Rewrite}\!\left(q^{(i)}, a^{(i)}, \mathcal{E}^{(i)}, \mathrm{type}^{(i)}\right),
\label{eq:rewrite_simple}
\end{equation}
where $\mathrm{Rewrite}(\cdot)$ is implemented via prompting the language model.
In practice, the rewrite sharpens entity/relation expressions when evidence is
off-topic, requests stricter evidence-grounded answering when unsupported claims
are detected, and adds missing constraints implied by the question when the
answer is incomplete.

\noindent\textbf{Bounded retry.}
The controller repeats retrieval and verification for at most
$I_{\max}\in\mathbb{N}$ iterations (typically small, e.g., $2$--$3$). If
$\mathrm{TripleCheck}(q^{(i)},a^{(i)},\mathcal{E}^{(i)})=1$ at any iteration,
the controller returns $(a^{(i)},\mathcal{E}^{(i)})$. Otherwise, it returns
\textsc{Fail}/\textsc{Abstain} after exhausting the budget.

\subsubsection{Optional: Human-in-the-Loop Knowledge Base Update}
\label{sec:hitl_update}

For domains where correctness and governance are critical, we optionally support
a human-in-the-loop (HITL) pathway to curate knowledge over time. After a query
is successfully verified, the system proposes candidate triples 
from the verified provenance $\mathcal{E}$ with source pointers. 
A human reviewer approves or rejects each proposal before insertion into the
knowledge graph. This module is orthogonal to the core control loop and is not
required for standard operation.

\subsection{Agentic Retriever}
\label{sec:agentic_retriever}

\begin{figure}[t]
  \centering
  \includegraphics[width=0.85\linewidth]{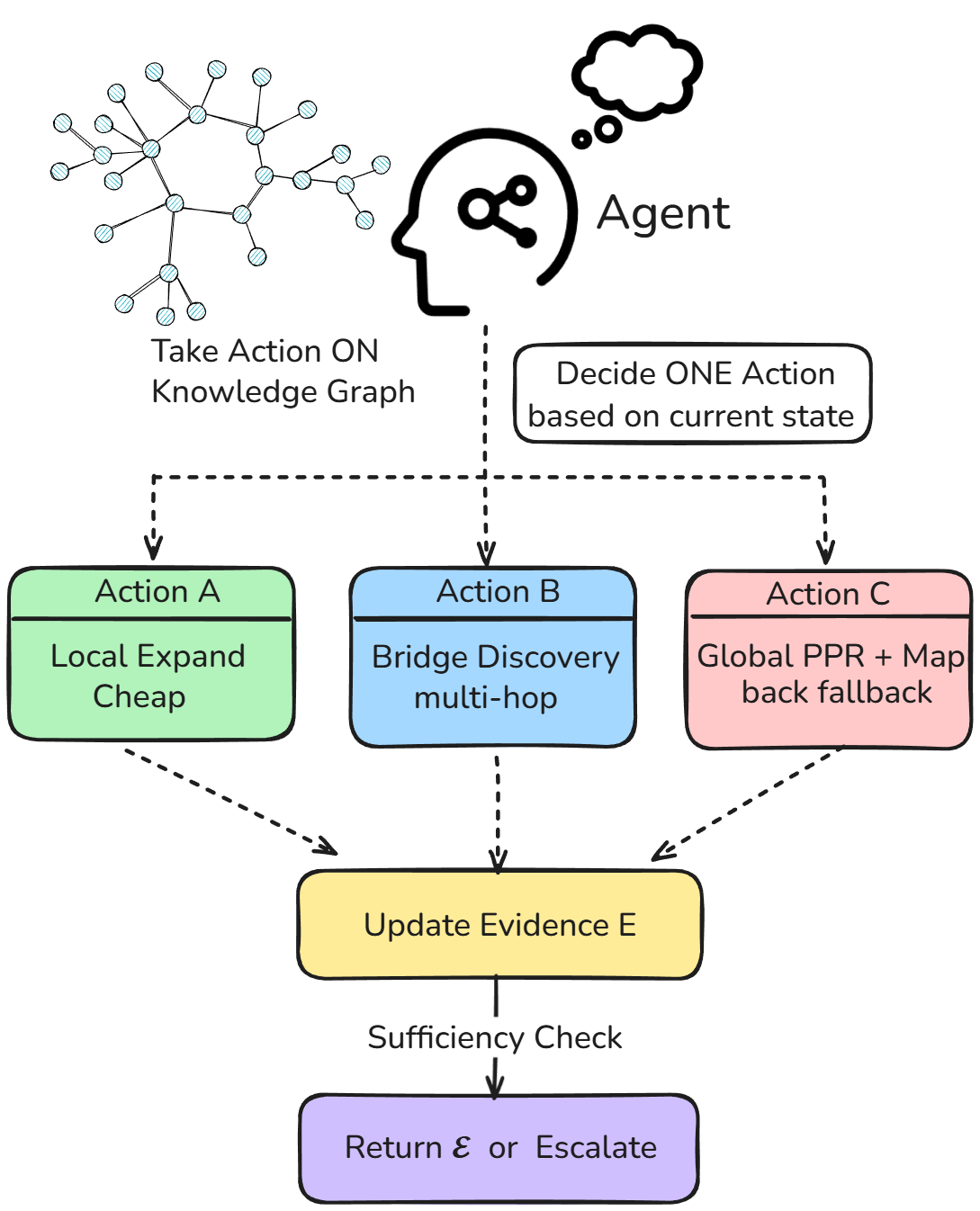}
  \caption{Agentic retriever}
  \label{fig:agentic_retriever}
  \vspace{-1.5em}
\end{figure}

The agentic retriever performs evidence acquisition once retrieval is invoked by
the adaptive control loop. Its design goal is to collect \emph{sufficient}
evidence at minimal cost while remaining robust to extraction loss in the
offline knowledge graph. To this end, the retriever operates as a stateful agent
that follows a local-first, escalation-on-demand policy: it begins with
inexpensive local graph operations and escalates to more global actions only when
the accumulated evidence is judged insufficient for answering. Importantly, the
retriever performs \emph{stage-wise evidence sufficiency checks} to guide
escalation, rather than certifying answer correctness.Unlike general-purpose tool-using agents that freely select actions, the retriever in A2RAG is an agent with an explicitly constrained action space and a monotonic escalation policy. This design trades expressive freedom for predictability, efficiency, and verifiable termination, which are critical in retrieval-centric systems.Figure~\ref{fig:agentic_retriever} provides an overview of the retriever state,
the local-first escalation hierarchy, and the provenance map-back fallback.

\noindent\textbf{Retriever state.}
At each step, the retriever maintains a state
$\langle q, \mathcal{S}_V, \mathcal{S}_R, \mathcal{E} \rangle$, where $q$ is the
current query (possibly rewritten by the controller), $\mathcal{S}_V$ denotes
aligned entity seeds, $\mathcal{S}_R$ denotes aligned relation seeds (optional),
and $\mathcal{E}$ is the accumulated evidence. Evidence is graph-structured
(triples) in Stages~1--2 and provenance text chunks
$\mathcal{E}\subseteq\mathcal{D}$ in Stage~3. The state is carried across stages,
and evidence is accumulated.

\noindent\textbf{Seed extraction and alignment.}
Given a query $q$, we extract entity mentions
$\widehat{\mathcal{E}}(q)=\{\hat{e}_1,\ldots,\hat{e}_m\}$ using a standard
NER/phrase extractor and align them to KG nodes via a hybrid lexical--semantic
matcher (e.g., edit similarity combined with embedding cosine), retaining only
high-confidence matches as $\mathcal{S}_V\subseteq\mathcal{V}$. Optionally, we
extract relation phrases $\widehat{\mathcal{R}}(q)=\{\hat{r}_1,\ldots,\hat{r}_n\}$
and align them to KG relation types to obtain $\mathcal{S}_R\subseteq\mathcal{R}$,
which is used as a lightweight filter to suppress edges inconsistent with the
query's relational intent.

\subsubsection{Stage 1: Local Evidence Collection}
\label{sec:local_retrieval}

Many practical queries can be answered from facts within a small neighborhood of
salient entities. Stage~1 therefore performs local expansion around entity seeds,
which is computationally cheap and often sufficient.

\noindent\textbf{Local neighborhood expansion.}
Given a knowledge graph $\mathcal{G}=(\mathcal{V},\mathcal{E}_{\mathcal{G}})$, the
1-hop neighbors of a seed $v\in\mathcal{S}_V$ are defined as
\begin{equation}
\mathcal{N}_1(v)
:= \{u\in\mathcal{V}\mid (v,r,u)\in\mathcal{E}_{\mathcal{G}}
\ \text{or}\ (u,r,v)\in\mathcal{E}_{\mathcal{G}},\ \exists r\in\mathcal{R}\}.
\label{eq:n1}
\end{equation}
If relation seeds $\mathcal{S}_R$ are available, we retain only edges whose
relation types are consistent with $\mathcal{S}_R$; otherwise, all incident
edges are kept. The union of 1-hop neighborhoods over $v\in\mathcal{S}_V$ forms a
small induced subgraph, which is serialized as graph-structured evidence.

\noindent\textbf{Rationale.}
Local-first retrieval targets the common case where relevant evidence is
concentrated near a few aligned entities. Early termination at this
stage avoids unnecessary global traversal and reduces exposure to noisy or
spurious paths.

\subsubsection{Stage 2: Bridge Discovery}
\label{sec:bridge_discovery}

When Stage~1 fails to connect multiple query entities, the missing evidence often
lies on short multi-hop connectors. Stage~2 searches for compact \emph{bridge
nodes} that jointly link multiple entity seeds, yielding a small but
high-signal subgraph.

\noindent\textbf{Bridge definition.}
To stabilize connectivity, we operate on an augmented graph $\mathcal{G}'$ that
includes inverse edges. Let $\mathcal{N}_K(v)$ denote the set of nodes within $K$
hops of $v$ in $\mathcal{G}'$, where $K\ge 2$ is a small hop budget. Bridge
candidates are defined as
\begin{equation}
\mathcal{B}_K :=
\Big\{u\in\mathcal{V}\ \Big|\ \big|\{v\in\mathcal{S}_V: u\in\mathcal{N}_K(v)\}\big|\ge 2 \Big\}.
\label{eq:bridge}
\end{equation}
This relaxed multi-seed constraint avoids brittle full intersections while
enforcing meaningful cross-entity connectivity.

\noindent\textbf{Bridge evidence construction.}
For each bridge node $b\in\mathcal{B}_K$, we extract a small number of short paths
(e.g., shortest paths) connecting $b$ to nearby seeds. The union of their triples
constitutes Stage~2 graph evidence. Both hop length and the number of paths per
pair are capped to control evidence size. If the resulting evidence remains
insufficient, the retriever escalates to Stage~3.

\subsubsection{Stage 3: Global Fallback with Degree-Normalized PPR and Provenance Map-back}
\label{sec:global_fallback}

Stages~1--2 may fail when the graph is sparse or suffers from extraction loss,
where fine-grained details (e.g., numerical values, temporal qualifiers,
exceptions) are absent from extracted triples. Stage~3 performs a global,
structure-guided retrieval using Personalized PageRank (PPR), followed by a
map-back to provenance text to recover such details.

\noindent\textbf{PPR formulation.}
Let $\mathcal{G}'=(\mathcal{V},\mathcal{E}')$ be the augmented graph and
$n:=|\mathcal{V}|$. Define the adjacency matrix $A\in\{0,1\}^{n\times n}$ with
$A_{uv}=1$ iff $(u,\cdot,v)\in\mathcal{E}'$, and the transition matrix
\begin{equation}
P_{uv} := \frac{A_{uv}}{\deg(u)},
\label{eq:trans}
\end{equation}
where $\deg(u)=\sum_v A_{uv}$. We construct a degree-normalized personalization
distribution for entity seeds $\mathcal{S}_V$:
\begin{equation}
p_0(u) :=
\begin{cases}
\frac{\deg(u)^{-1}}{Z}, & u\in\mathcal{S}_V,\\
0, & \text{otherwise},
\end{cases}
\qquad
Z:=\sum_{u\in\mathcal{S}_V}\deg(u)^{-1}.
\label{eq:p0}
\end{equation}
With teleport probability $\alpha\in(0,1)$, the PPR score vector $\mathbf{r}$ is
defined by the fixed point
\begin{equation}
\mathbf{r} = \alpha \mathbf{p}_0 + (1-\alpha) P^\top \mathbf{r}.
\label{eq:ppr}
\end{equation}

\noindent\textbf{Provenance map-back.}
We select the top-$L$ nodes by PPR score and map them back to provenance text
chunks using an offline mapping $\pi:\mathcal{V}\to 2^{\mathcal{D}}$. The final
text evidence is
\begin{equation}
\mathcal{E} := \bigcup_{u\in \mathcal{V}_{\mathrm{top}\text{-}L}} \pi(u)
\subseteq \mathcal{D}.
\label{eq:mapback}
\end{equation}
This map-back step is essential for robustness to extraction loss: it leverages
global graph structure for routing while grounding the final answer in the
original source text.

\subsubsection{Escalation Policy and Termination}
\label{sec:escalation_policy}

The retriever follows a monotonic escalation hierarchy
(Local $\rightarrow$ Bridge $\rightarrow$ Global). Each stage is invoked at most
once per query, and evidence sufficiency is evaluated after each stage to decide
whether escalation is necessary. Since the number of stages is finite and each
stage is bounded by explicit budgets, the retrieval process always terminates.

\subsection{End-to-End Summary}
\label{sec:e2e_summary}

\begin{algorithm}[t]
\caption{A2RAG End-to-End Procedure}
\label{alg:a2rag}
\begin{algorithmic}[1]
\Require Query $q$, corpus $\mathcal{D}$, summaries $\mathcal{S}=\{s_j\}_{j=1}^{N}$, KG $\mathcal{G}=(\mathcal{V},\mathcal{E}_{\mathcal{G}})$, threshold $\tau_g$, max retries $I_{\max}$
\Ensure Verified answer $a$ with evidence $\mathcal{E}\subseteq\mathcal{D}$, or \textsc{Abstain}/\textsc{Fail}

\State \textbf{// Entry: summarized-KB gating}
\State $q^{(0)} \gets q$
\If{$\mathrm{Gate}(q^{(0)})=0$}
    \State \Return \textsc{Abstain} \Comment{out-of-scope or insufficient corpus coverage}
\EndIf

\For{$i \gets 0$ to $I_{\max}$}

    \State \textbf{// Evidence acquisition (agentic retriever)}
    \State $\mathcal{E}^{(i)} \gets \textsc{AgenticRetrieve}(q^{(i)}, \mathcal{G}, \mathcal{D})$
    \Comment{local-first, escalation-on-demand retrieval}

    \State \textbf{// Candidate answering}
    \State $a^{(i)} \gets \textsc{GenerateAnswer}(q^{(i)}, \mathcal{E}^{(i)})$

    \State \textbf{// Exit: answer-level verification}
    \If{$\mathrm{TripleCheck}(q^{(i)}, a^{(i)}, \mathcal{E}^{(i)})=1$}
        \State $a \gets a^{(i)};\ \mathcal{E}\gets \mathcal{E}^{(i)}$
        \State \textsc{OptionalHITLUpdate}$(\mathcal{E}, \mathcal{G})$
        \State \Return $a$ \Comment{verified and grounded}
    \EndIf

    \State \textbf{// Iteration: failure-aware rewriting}
    \State $\mathrm{type}^{(i)} \gets \textsc{FailureType}(q^{(i)}, a^{(i)}, \mathcal{E}^{(i)})$
    \State $q^{(i+1)} \gets \mathrm{Rewrite}\!\left(q^{(i)}, a^{(i)}, \mathcal{E}^{(i)}, \mathrm{type}^{(i)}\right)$

\EndFor

\State \Return \textsc{Fail} \Comment{retry budget exhausted}
\end{algorithmic}
\end{algorithm}

Algorithm~\ref{alg:a2rag} summarizes the end-to-end execution of A2RAG under a
clear separation of responsibilities between \emph{global control} and
\emph{agentic retrieval}. The controller applies summarized-KB gating to filter
out out-of-scope queries, verifies candidate answers against provenance evidence
via Triple-Check, and performs bounded, failure-aware query rewriting when
necessary. Once invoked, the agentic retriever operates as a stateful agent with
a constrained action space and a monotonic, local-first escalation policy to
collect sufficient evidence at minimal cost. The procedure always terminates
within a bounded budget, returning a verified and grounded answer when possible,
or abstaining when reliable evidence cannot be obtained.

\section{Experiments}
\label{sec:eval}

\subsection{Experimental Setup}
\label{sec:setup}

\subsubsection{Tasks and Datasets}
We evaluate A2RAG on multi-hop, knowledge-intensive question answering (QA),
where answering requires integrating evidence distributed across multiple
documents. Experiments are conducted on \textbf{HotpotQA}~\cite{yang2018hotpotqa}
and \textbf{2WikiMultiHopQA}~\cite{ho2020constructing}.

\noindent\textbf{Dataset setting.}
Due to resource constraints, we run all experiments on \emph{subsets} of the
above benchmarks by sampling a fixed number of instances from each dataset.
This setting is sufficient for controlled comparison, but scaling to the full
datasets is left for future work.

\subsubsection{Baselines and Evaluation Protocol}
Our evaluation follows two protocols.

\textbf{(i) General QA performance.}
We compare A2RAG against three representative modes:
\textbf{NoRAG} (base LLM without retrieval), \textbf{TextRAG} (dense passage
retrieval over the corpus), and \textbf{LightRAG}~\cite{guo2025lightrag}
(local graph-based retrieval with neighborhood-level context).
We report end-task QA performance (\textbf{EM}/\textbf{F1}) and retrieval
quality (\textbf{Recall@$K$}, $K\in\{2,5\}$).

\textbf{(ii) Multi-hop efficiency.}
To assess the cost of multi-step retrieval, we compare A2RAG with
\textbf{IRCoT}~\cite{trivedi2023ircot}, an iterative retrieval baseline designed
for multi-hop evidence accumulation. We report \textbf{latency} and \textbf{cost}
metrics (token usage and/or number of LLM calls) and retrieval/QA
quality.

\subsubsection{Models}
Unless otherwise specified, the backbone LLM is \textbf{gpt-4o-mini} with
deterministic decoding (temperature $=0$). For dense retrieval components, we
use \textbf{text-embedding-3-small} as the retrieval encoder. All baselines are
configured to use the same backbone models to ensure a fair comparison.

\subsubsection{A2RAG-Specific Measurements}
To characterize the behavior of progressive retrieval, we additionally report:
(i) the fraction of queries resolved by \textbf{local 1-hop retrieval}, by
\textbf{$K$-hop bridge discovery}, and by the \textbf{PPR-based global fallback},
(ii) an ablation that removes \textbf{relation seeding} (node-only seeds) to
quantify the benefit of relation-aware retrieval, (iii) a focused comparison of
retrieved provenance chunks for \textbf{PPR map-back} versus \textbf{TextRAG}
on instances where PPR is triggered, and (iv) a robustness test that simulates
\textbf{extraction loss} by deleting nodes/edges from the KG and evaluating
performance degradation.

\noindent\textbf{Evaluation map.}
We evaluate by addressing three core questions:
\emph{(Q1)} Does A2RAG improve evidence retrieval under small-$K$ budgets?
(Table~\ref{tab:main_results});
\emph{(Q2)} Does progressive escalation reduce multi-hop cost compared to an
iterative retrieve--reason baseline? (Tables~\ref{tab:eff_hotpot}
and~\ref{tab:eff_2wiki}, Fig.~\ref{fig:stage_breakdown});
\emph{(Q3)} Is A2RAG robust to extraction loss, and does PPR map-back recover
higher-quality provenance evidence? (Figs.~\ref{fig:robust_hotpot}
and~\ref{fig:robust_2wiki}, Sec.~\ref{sec:robust_extraction}).

\subsection{Main Results on Multi-hop QA Benchmarks}
\label{sec:main_results}

Table~\ref{tab:main_results} summarizes the results on HotpotQA~\cite{yang2018hotpotqa}
and 2WikiMultiHopQA~\cite{ho2020constructing}. Overall, A2RAG achieves the
strongest evidence retrieval performance on both benchmarks. On HotpotQA, A2RAG
reaches Recall@2/Recall@5 of 62.4/73.6, exceeding the strongest LightRAG mode
(mix)~\cite{guo2025lightrag} (56.8/67.5). On 2WikiMultiHopQA, A2RAG again yields
the best recall (58.9/69.2), outperforming LightRAG (mix) (52.7/63.8). These
gains validate the central design choice of A2RAG: a progressive retrieval
policy that first attempts compact mid-range bridge discovery and only activates
PPR diffusion as a last-resort fallback, while always grounding retrieval in
provenance chunks. 

In terms of answer accuracy, A2RAG is competitive but does not always attain
the best EM/F1. On HotpotQA, LightRAG (mix) achieves the highest EM/F1
(33.9/46.5), while A2RAG obtains 32.2/43.7. On 2WikiMultiHopQA, LightRAG (mix)
achieves the best EM (31.5) and A2RAG achieves comparable F1 (42.9 vs.\ 41.5),
but slightly lower EM (30.0 vs.\ 31.5). This pattern is consistent with the
objective and control mechanism of A2RAG: we explicitly optimize the retriever
for high-confidence evidence inclusion under small-$K$ budgets (reflected by the
largest Recall@K gains), and the bounded verification loop prioritizes strict
groundedness and adequacy, which can trade off some generation flexibility and
surface-form matching measured by EM/F1. We further contextualize this trade-off
via system-level efficiency (Sec.~\ref{sec:efficiency_ircot}) and mechanistic
analysis of progressive escalation (Sec.~\ref{sec:breakdown}).

\noindent\textbf{Production-level Dataset.} 
We observe a similar trend in a real-world QA setting on a production-level dataset provided by our industry partner~\cite{wang2025aefa}, based on financial trading platform operation manuals. On this dataset, A2RAG improves end-to-end Recall@5 by approximately 15\% over LightRAG (mix) and remains more robust under incomplete KGs, retaining substantially higher Recall@5 under graph degradation (e.g., $67.7$ vs.\ $46.5$ at $20\%$ KG node/edge removal).

\begin{table}[t]
\centering
\caption{Results on HotpotQA~\cite{yang2018hotpotqa} and
2WikiMultiHopQA~\cite{ho2020constructing} (200 questions per dataset).
We report EM/F1 and evidence-level Recall@K (K=2,5).}
\label{tab:main_results}
\small
\setlength{\tabcolsep}{3.8pt}
\begin{tabular}{l|cc|cc}
\toprule
\textbf{Method} & \textbf{EM} $\uparrow$ & \textbf{F1} $\uparrow$ & \textbf{R@2} $\uparrow$ & \textbf{R@5} $\uparrow$ \\
\midrule
\multicolumn{5}{c}{\textbf{HotpotQA}} \\
\midrule
No RAG & 21.3 & 29.8 & -- & -- \\
Naive RAG & 28.4 & 37.6 & 44.0 & 59.0 \\
LightRAG (local)~\cite{guo2025lightrag} & 32.4 & 43.2 & 52.0 & 65.2 \\
LightRAG (global)~\cite{guo2025lightrag} & 31.6 & 42.5 & 53.1 & 66.0 \\
LightRAG (mix)~\cite{guo2025lightrag} & \textbf{33.9} & \textbf{46.5} & 56.8 & 67.5 \\
A2RAG (Ours) & 32.2 & 43.7 & \textbf{62.4} & \textbf{73.6} \\
\midrule
\multicolumn{5}{c}{\textbf{2WikiMultiHopQA}} \\
\midrule
No RAG & 17.8 & 25.7 & -- & -- \\
Naive RAG & 24.9 & 34.2 & 40.8 & 54.2 \\
LightRAG (local)~\cite{guo2025lightrag} & 28.9 & 39.6 & 48.6 & 60.7 \\
LightRAG (global)~\cite{guo2025lightrag} & 29.4 & 40.0 & 50.4 & 62.3 \\
LightRAG (mix)~\cite{guo2025lightrag} & \textbf{31.5} & 41.5 & 52.7 & 63.8 \\
A2RAG (Ours) & 30.0 & \textbf{42.9} & \textbf{58.9} & \textbf{69.2} \\
\bottomrule
\end{tabular}
\end{table}

\subsection{Efficiency on Multi-hop Queries: A2RAG vs.\ IRCoT}
\label{sec:efficiency_ircot}

IRCoT~\cite{trivedi2023ircot} is a strong iterative baseline for multi-hop
question answering, but its retrieve--reason loops can incur substantial runtime
overhead due to repeated LLM invocations and prompt growth across steps. We
therefore evaluate the efficiency of A2RAG against IRCoT on multi-hop queries,
focusing on system-level cost and latency under a fair and controlled setting.

\noindent\textbf{Multi-hop queries.}
We construct a multi-hop subset from the sampled evaluation data by selecting
instances whose supporting facts span multiple documents/pages. This yields 180
queries on HotpotQA and 150 queries on 2WikiMultiHopQA.

\noindent\textbf{Fair comparison protocol.}
We align the maximum number of iterations by constraining IRCoT to a fixed step
limit and A2RAG to bounded rewrite--retry rounds, and keep the retrieved
evidence budget comparable across methods.

\noindent\textbf{Metrics.}
We report (i) total token usage (prompt + completion), (ii) the number of LLM
calls per query, and (iii) end-to-end latency. For latency, we report both mean
and tail latency (P95) to reflect serving-time stability.

\noindent\textbf{Results.}
Tables~\ref{tab:eff_hotpot} and~\ref{tab:eff_2wiki} show that A2RAG is
substantially more efficient than IRCoT on multi-hop queries. On HotpotQA (MH),
A2RAG reduces total token usage from 30K to 16K and decreases the average number
of LLM calls from 3.5 to 2.0. It also improves latency, achieving 2.7s mean
latency (P95 4.2s) compared to 4.8s (P95 7.5s) for IRCoT. Similar gains are
observed on 2WikiMultiHopQA (MH), where A2RAG lowers token consumption from 35K
to 18K, reduces LLM calls from 4.2 to 2.3, and decreases mean latency from 5.6s
(P95 8.8s) to 3.2s (P95 5.0s).

\noindent\textbf{Why A2RAG is faster.}
The efficiency advantage stems from A2RAG's progressive retrieval policy. In
contrast, A2RAG resolves multi-hop connections via bounded bridge
discovery, and when local and bridge evidence are insufficient it performs a
\emph{single} structure-aware diffusion step (PPR) followed by provenance
map-back, avoiding repeated iterative prompting. This reduces both the number
of iterations and the per-iteration context growth, leading to lower token usage
and latency.

\begin{table}[t]
\centering
\caption{Efficiency on HotpotQA (MH), $N=180$. Lower is better.}
\label{tab:eff_hotpot}
\small
\setlength{\tabcolsep}{4.5pt}
\begin{tabular}{l|r r r r}
\toprule
\textbf{Method} & \textbf{Tokens} $\downarrow$ & \textbf{Calls} $\downarrow$ &
\textbf{Lat} $\downarrow$ & \textbf{P95} $\downarrow$ \\
\midrule
IRCoT~\cite{trivedi2023ircot} & 30k & 3.5 & 4.8 & 7.5 \\
A2RAG & 16k & 2.0 & 2.7 & 4.2 \\
\bottomrule
\end{tabular}
\end{table}

\begin{table}[t]
\centering
\caption{Efficiency on 2WikiMultiHopQA (MH), $N=150$. Lower is better.}
\label{tab:eff_2wiki}
\small
\setlength{\tabcolsep}{4.5pt}
\begin{tabular}{l|r r r r}
\toprule
\textbf{Method} & \textbf{Tokens} $\downarrow$ & \textbf{Calls} $\downarrow$ &
\textbf{Lat} $\downarrow$ & \textbf{P95} $\downarrow$ \\
\midrule
IRCoT~\cite{trivedi2023ircot} & 35k & 4.2 & 5.6 & 8.8 \\
A2RAG & 18k & 2.3 & 3.2 & 5.0 \\
\bottomrule
\end{tabular}
\end{table}

\subsection{Progressive Retrieval Breakdown}
\label{sec:breakdown}

\begin{figure}[t]
  \centering
  \includegraphics[width=\columnwidth]{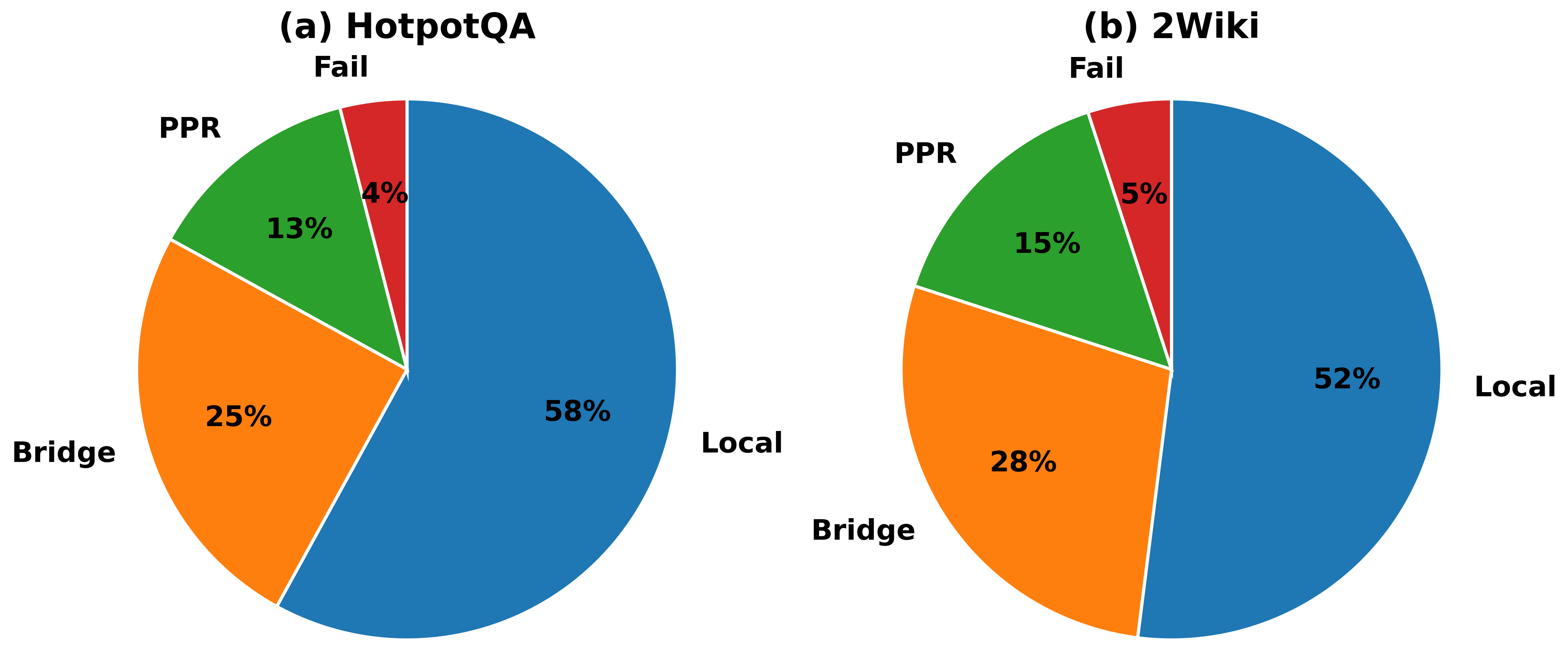}
  \caption{Stage-wise breakdown of A2RAG's progressive retrieval. Each pie
  chart reports the fraction of queries that terminate at the local (1-hop),
  bridge ($K$-hop), or PPR-based global fallback stage, with failed cases shown
  separately.}
  \label{fig:stage_breakdown}
\end{figure}

We analyze how A2RAG allocates retrieval effort across stages by recording the
stage at which a query terminates (Local 1-hop, $K$-hop Bridge, or PPR-based
global fallback), and reporting failed cases separately. As shown in
Fig.~\ref{fig:stage_breakdown}, most queries are resolved without invoking the
global fallback. On HotpotQA, 58\% of queries terminate at the local 1-hop stage
and 25\% terminate after bridge discovery, while only 13\% require the PPR-based
fallback; 4\% fail after bounded retries. A similar pattern holds on
2WikiMultiHopQA, where 52\% terminate locally and 28\% terminate at the bridge
stage, with 15\% requiring PPR and 5\% failing.

This distribution provides a mechanistic explanation for the efficiency gains
reported in Sec.~\ref{sec:efficiency_ircot}. A2RAG reserves the most expensive
operation (global diffusion with provenance map-back) for a minority of hard
cases, while the majority are handled by inexpensive local evidence collection
or bounded mid-range bridge reasoning. Meanwhile, the non-trivial fraction of
PPR-triggered queries suggests that leveraging KG structure for global diffusion
can be essential for recovering distributed evidence and mitigating extraction
loss when local or bridge-level graph evidence is insufficient, which we study
next in Sec.~\ref{sec:robust_extraction}.

\subsection{Ablation: The Role of Relation Seeding}
\label{sec:ablation_relation}

A2RAG extracts and aligns both entity seeds $S_V$ and relation-intent seeds
$S_R$ from the query to steer evidence collection toward the relations implied
by the question. To quantify the contribution of relation seeding, we perform
an ablation that disables $S_R$ and keeps all other components unchanged
(backbone LLM, dense encoder, progressive retrieval stages, and verification
loop). The resulting \textbf{node-only} variant relies solely on entity seeds
for local neighborhood retrieval, bridge discovery, and PPR map-back.

Table~\ref{tab:ablation_relation_seed} shows that removing relation seeds
consistently degrades both evidence recall and end-task QA accuracy. On HotpotQA,
the node-only variant drops from 62.4/73.6 to 56.1/69.8 in Recall@2/Recall@5,
and from 32.2/43.7 to 29.8/40.5 in EM/F1. On 2WikiMultiHopQA, Recall@2/Recall@5
decreases from 58.9/69.2 to 51.5/64.7, with EM/F1 dropping from 30.0/42.9 to
27.2/38.7. Notably, the recall reduction is more pronounced at small $K$
(Recall@2), indicating that relation seeding improves retrieval
\emph{directionality} and helps surface high-signal evidence early, rather than
relying on broader context expansion.

These results support the motivation behind incorporating $S_R$ in A2RAG.
Entity-only seeding often yields ambiguous local neighborhoods where many edges
are unrelated to the query's relational intent. Relation seeds provide an
additional constraint that focuses both local evidence collection and bridge
search on relation-consistent connectors, thereby reducing topological drift and
improving the quality of the evidence set under the same retrieval budget.

\begin{table}[t]
\centering
\caption{Ablation on relation seeding. ``Full'' uses entity and relation seeds
($S_V+S_R$), while ``Node-only'' disables relation seeds ($S_V$ only).}
\label{tab:ablation_relation_seed}
\small
\setlength{\tabcolsep}{4.2pt}
\begin{tabular}{l|l|c c|c c}
\toprule
\textbf{Dataset} & \textbf{Variant} & \textbf{EM} & \textbf{F1} & \textbf{R@2} & \textbf{R@5} \\
\midrule
\multirow{2}{*}{HotpotQA} 
& Full ($S_V+S_R$) & 32.2 & 43.7 & 62.4 & 73.6 \\
& Node-only ($S_V$) & 29.8 & 40.5 & 56.1 & 69.8 \\
\midrule
\multirow{2}{*}{2WikiMultiHopQA} 
& Full ($S_V+S_R$) & 30.0 & 42.9 & 58.9 & 69.2 \\
& Node-only ($S_V$) & 27.2 & 38.7 & 51.5 & 64.7 \\
\bottomrule
\end{tabular}
\end{table}

\subsection{Robustness to Extraction Loss}
\label{sec:robust_extraction}

A central practical challenge in graph-based RAG is \emph{extraction loss}:
the constructed knowledge graph inevitably misses nodes, relations, or
fine-grained attributes present in the raw text, and the graph may further
degrade under incremental updates or imperfect extraction pipelines. To evaluate
robustness under such lossy conditions, we simulate extraction loss by randomly
deleting a fixed percentage of nodes and their incident edges from the KG while
keeping the underlying text corpus unchanged. We then measure retrieval quality
using \textbf{Recall@5} on HotpotQA and 2WikiMultiHopQA, comparing A2RAG against
a graph-only baseline (LightRAG~\cite{guo2025lightrag}) and a text-only baseline
(TextRAG).

\subsubsection{Deletion Stress Test}
\begin{figure}[t]
  \centering
  \includegraphics[width=0.8\columnwidth]{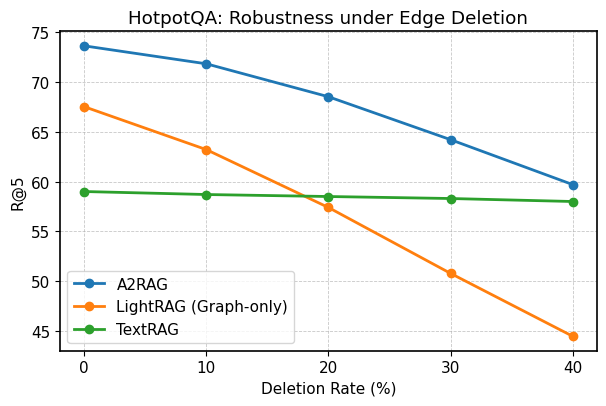}
  \caption{Robustness to extraction loss on HotpotQA measured by Recall@5 under
  random KG node/edge deletion.}
  \label{fig:robust_hotpot}
  \vspace{-1em}
\end{figure}

\begin{figure}[t]
  \centering
  \includegraphics[width=0.8\columnwidth]{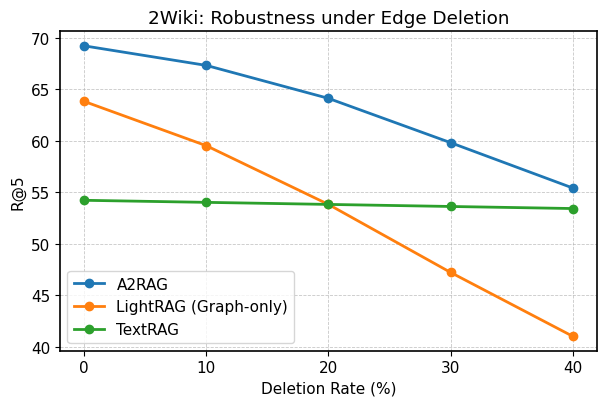}
  \caption{Robustness to extraction loss on 2WikiMultiHopQA measured by Recall@5
  under random KG node/edge deletion.}
  \label{fig:robust_2wiki}
  \vspace{-1em}
\end{figure}

Figure~\ref{fig:robust_hotpot} and Figure~\ref{fig:robust_2wiki} show that A2RAG
degrades substantially more gracefully than the graph-only baseline as deletion
increases. On HotpotQA, LightRAG drops from 67.5 (0\%) to 44.5 (40\%), whereas
A2RAG decreases from 73.6 to 59.7 over the same range. A similar pattern holds
on 2WikiMultiHopQA, where LightRAG falls from 63.8 to 41.0, while A2RAG declines
from 69.2 to 55.4. In contrast, TextRAG remains nearly unchanged across deletion
levels (e.g., 59.0$\rightarrow$58.0 on HotpotQA and 54.2$\rightarrow$53.4 on
2WikiMultiHopQA), consistent with the fact that the text corpus is not modified.

These results support the key design of A2RAG: although graph structure is used
for efficient navigation and multi-hop connectivity, the system ultimately
grounds evidence in provenance chunks from the original text via the PPR map-back
mechanism. When the KG is partially missing, graph-only retrieval suffers
sharply because local neighborhoods and bridge connectors become unreliable. In
contrast, A2RAG can still leverage the remaining structural signals to locate
high-relevance regions and recover fine-grained evidence from source chunks,
thereby maintaining higher recall under increasing extraction loss. At high
deletion ratios, A2RAG gradually approaches the text-only baseline, which is
expected when the graph becomes insufficient to provide additional structural
guidance beyond flat retrieval.

\subsubsection{Effectiveness of PPR-based Provenance Map-back}
\label{sec:ppr_vs_text}

To further understand the source of A2RAG’s robustness under extraction loss, we
compare the effectiveness of provenance chunks retrieved via the PPR map-back
mechanism with those retrieved by standard text-only retrieval. We observe a
consistent pattern across datasets: chunks selected through PPR-guided mapping
are more likely to contain the critical facts required to answer the query,
compared to top-ranked chunks retrieved by TextRAG.

This advantage stems from the structural bias introduced by graph diffusion.
Rather than relying solely on surface-level lexical similarity, PPR leverages
the global connectivity of the knowledge graph to identify nodes that jointly
explain multiple query seeds. Mapping these high-confidence structural anchors
back to source chunks yields evidence that is both topically relevant and
structurally grounded, often surfacing rare but decisive details such as numeric
constraints or temporal qualifiers -- that flat text retrieval frequently overlooks.

As a result, even when the knowledge graph is incomplete, the PPR-based map-back
mechanism enables A2RAG to recover higher-quality evidence from the original
corpus. This observation aligns with the robustness trends reported above, where
A2RAG consistently outperforms both graph-only and text-only baselines under
increasing extraction loss.

\subsection{Discussion and Limitations}
\label{sec:discussion}

\noindent\textbf{Evaluation Scale and Generalization.}
Due to the cost of controlled multi-hop evaluation, our experiments are
conducted on subsets of HotpotQA~\cite{yang2018hotpotqa} and
2WikiMultiHopQA~\cite{ho2020constructing}. While the results are consistent
across datasets and experimental settings, larger-scale evaluations on more
diverse corpora would further strengthen the generality of our conclusions.

\noindent\textbf{Sensitivity to Seed Quality.}
The effectiveness of progressive retrieval depends on the quality of entity and
relation seed extraction. Inaccurate or incomplete seeding can lead to
suboptimal local neighborhoods or misdirected bridge discovery, increasing
reliance on global fallback. Although the verification and query rewriting loop
mitigates some of these failures, improving robust and domain-adaptive seed
extraction remains an important direction for future work.

\noindent\textbf{Scope of Structural Guidance and Maintenance.}
A2RAG relies on the knowledge graph primarily as a structural index rather than
a complete semantic store. When the graph becomes extremely sparse or severely
fragmented, the benefit of structure-guided retrieval naturally diminishes, and
the system behavior gradually approaches that of text-only retrieval. While this
degradation is graceful (Sec.~\ref{sec:robust_extraction}), it highlights that
A2RAG still assumes the presence of a minimally informative graph backbone to
provide effective navigation signals. In addition, studying A2RAG under
continuously evolving knowledge bases and long-term updates remains an open
challenge, particularly when extraction pipelines introduce drift or partial
graph degradation over time.

Overall, these limitations do not detract from the core contribution of A2RAG
but instead clarify the conditions under which adaptive, agentic graph retrieval
is most effective, and point toward promising avenues for extending the
framework.

\section{Conclusion}
\label{sec:conclusion}

We introduced A2RAG, an adaptive and agentic GraphRAG framework that addresses two practical bottlenecks in existing systems: one-size-fits-all retrieval under mixed-difficulty workloads and vulnerability to extraction loss in imperfect knowledge graphs. Rather than treating graph retrieval as a static pipeline, A2RAG formulates retrieval as a controlled, cost-aware process that progressively acquires evidence with stage-wise sufficiency checks, combining local-first expansion, bounded bridge discovery, and a structure-guided PPR fallback with provenance map-back to recover fine-grained, verifiable text evidence. Experiments on multi-hop QA benchmarks show that A2RAG improves retrieval effectiveness and efficiency over strong text-based and graph-based baselines while remaining resilient to incomplete and lossy graph construction, making graph-augmented LLM question answering more reliable and practical under realistic deployment constraints.


\bibliographystyle{IEEEtran}
\bibliography{references}

\end{document}